\documentclass[aps,prd,twocolumn,preprintnumbers,superscriptaddress]{revtex4-1}
\usepackage{graphicx}
\usepackage{epstopdf}
\usepackage{amsmath}
\usepackage{amsfonts}
\usepackage{amssymb}
\usepackage{appendix}
\usepackage{enumerate}
\usepackage{natbib}
\usepackage{comment}
\usepackage{bbold}
\usepackage[shortlabels]{enumitem}
\usepackage{color}
\usepackage{slashed}
\usepackage{subfigure}
\usepackage{setspace}
\usepackage{footnote}
\usepackage{lipsum}
\usepackage{multirow}
\usepackage[colorlinks = true,
            linkcolor = blue,
            urlcolor  = blue,
            citecolor = blue,
            anchorcolor = blue]{hyperref}
\usepackage[capitalize]{cleveref}
\usepackage{braket}
\usepackage[compat=1.1.0]{tikz-feynman}
\usepackage{multirow}
\usepackage{physics}
\usepackage{feynmp-auto}
\usepackage[normalem]{ulem}
\usepackage{url}
\usepackage{units}
\usepackage[normalem]{ulem}
\usepackage{bm}

\newcommand{\be}{\begin{eqnarray}}
\newcommand{\ee}{\end{eqnarray}}
\newcommand{\ba} {\begin{equation}\begin{aligned}}
\newcommand{\ea} {\end{aligned}\end{equation}}
\newcommand{\bg} {\begin{equation}\begin{gathered}}
\newcommand{\eg} {\end{gathered}\end{equation}}

\newcommand{\beq}{\begin{equation}}
\newcommand{\eeq}{\end{equation}}

\newcommand{\sL}{\mathcal{L}}

\usepackage{tikz,xcolor,hyperref}

\definecolor{lime}{HTML}{A6CE39}
\DeclareRobustCommand{\orcidicon}{\hspace{-1mm}
	\begin{tikzpicture}
		\draw[lime, fill=lime] (0,0) 
		circle [radius=0.12] 
		node[white] {{\fontfamily{qag}\selectfont \tiny \,ID}};
		\draw[white, fill=white] (-0.0525,0.095) 
		circle [radius=0.007];
	\end{tikzpicture}
	\hspace{-3mm}
}

\foreach \x in {A, ..., Z}{\expandafter\xdef\csname orcid\x\endcsname{\noexpand\href{https://orcid.org/\csname orcidauthor\x\endcsname}
		{\noexpand\orcidicon}}
}

\begin{document}

\title{Neutron Star Eclipses as Axion Laboratories}
	\author{Vedran~Brdar\orcidA{}}
	\email{vedran.brdar@okstate.edu}
	\affiliation{Department of Physics, Oklahoma State University, Stillwater, OK 74078, USA}
	\author{Dibya~S.~Chattopadhyay\orcidB{}}
	\email{dibya.chattopadhyay@okstate.edu}
	\affiliation{Department of Physics, Oklahoma State University, Stillwater, OK 74078, USA}

\begin{abstract}
In light-shining-through-walls experiments, axions and axion-like particles (ALPs) are searched for by exposing an optically thick barrier to a laser beam. In a magnetic field, photons could convert into ALPs in front of the barrier and reconvert behind it, giving rise to a signal that can occur only in the presence of such hidden particles. In this work, we utilize the light-shining-through-walls concept and apply it to astrophysical scales. Namely, we consider eclipsing binary systems, consisting of a neutron star, which is a bright source of X-rays, and a companion star with a much larger radius. Space observatories such as XMM-Newton and NuSTAR have performed extensive measurements of such systems, obtaining data on both out-of-eclipse photon rates and those during eclipses. The latter are typically $\mathcal{O}(10^2-10^3)$ times smaller, due to the fact that X-rays propagating along the line of sight from the neutron star to the X-ray observatory do not pass through the barrier that is the companion star. Using this attenuation, we derive a constraint on  ALP-photon coupling of $g_{a\gamma} \leq 1.44 \times 10^{-10} \,\text{GeV}^{-1}$ (at 90\% C.L.) for the LMC X-4 eclipsing binary system, surpassing current bounds from light-shining-through-walls experiments. We also present future prospects that could realistically improve this limit by an order of magnitude in $g_{a\gamma}$, making it competitive with some of the strongest limits derived to date.
\end{abstract}

\maketitle

\noindent
\section{Introduction}
There are very few observations in high-energy physics that cannot be explained within the Standard Model (SM). One of them is the apparent conservation of CP symmetry in strong interactions, known as the strong CP problem \cite{Dine:2000cj}. The proposed solutions to this problem include spontaneous CP violation \cite{Nelson:1983zb,Barr:1984qx,Nelson:1984hg}, spontaneous parity violation \cite{Babu:1989rb}, and a massless up quark \cite{Banks:1994yg,Alexandrou:2020bkd}. However, by far the most widely studied explanation of the strong CP problem is the Peccei-Quinn mechanism \cite{Peccei:1977hh}, which features the spontaneous breaking of a $U(1)$ symmetry that gives rise to the QCD axion \cite{Weinberg:1977ma,Wilczek:1977pj}. Axion-like particles (ALPs) appear in many BSM theories and can also serve as dark matter candidates \cite{Preskill:1982cy,Abbott:1982af,Dine:1982ah,Adams:2022pbo}. They are analogous to QCD axions, with the main difference being that their mass and decay constant are independent.
 If such particles exist, they can interact with the SM through the effective ALP-photon operator~\cite{Raffelt:1987im}
\begin{align}
\sL_{a\gamma} = -\frac{1}{4}\,g_{a\gamma}\,a\,F^{\mu\nu}\,\tilde{F}_{\mu\nu}\,,
\label{eq:lag}
\end{align}
where $a$ is the ALP field, $F^{\mu\nu}$ is the field strength tensor of the electromagnetic field ($\tilde{F}_{\mu\nu}$ is its dual) and $g_{a\gamma}$ is the ALP-photon coupling strength. This interaction can be used to constrain ALPs through numerous cosmological \cite{Cadamuro:2011fd,Cadamuro:2010cz,Balazs:2022tjl}, astrophysical \cite{Caputo:2024oqc}, and terrestrial probes \cite{OSQAR:2007oyv,ALPS:2009des,Betz:2013dza,Ejlli:2020yhk}; for a compilation of available limits see \cite{AxionLimits}. 

The leading terrestrial probes are light-shining-through-walls experiments which are searching for ALPs by illuminating a barrier with laser beam. Given the interaction term in \cref{eq:lag}, photons can partially convert into ALPs before reaching the barrier and reconvert after it, leading to a signal that is only possible if such hidden particles exist. The absence of the signal has placed strong constraints on $g_{a\gamma}$ \cite{OSQAR:2007oyv,ALPS:2009des,Betz:2013dza,ALPSII:2025eri}.

The light-shining-through-walls concept has also been studied in astrophysical settings. Namely, this was done by one of us in Ref.~\cite{Chattopadhyay:2023nuq}, in the context of magnetars obscured by a nebula. However, the large uncertainties in the estimate of intrinsic luminosity of an obscured magnetar lead to a large uncertainty in the derived constraint. Further, see Ref.~\cite{Fairbairn:2006hv} for how ALPs can make the Sun partially transparent for high energy photons, and Ref.~\cite{Frerick:2022mjg} for an implementation of the light-shining-through-wall concept for solar dark photons, using data from a solar eclipse.

In this work, we apply the light-shining-through-walls principle to an eclipsing binary system composed of a neutron star and a larger main sequence star. Neutron stars are bright sources of X-rays, the flux of which is strongly attenuated during eclipses when the neutron star is blocked by its companion star. In this sense, eclipses in binary systems resemble light-shining-through-walls experiments, where the companion star serves as a barrier and the X-ray space observatories as the detectors.

\begin{figure*}
    \centering
    \includegraphics[width=\textwidth]{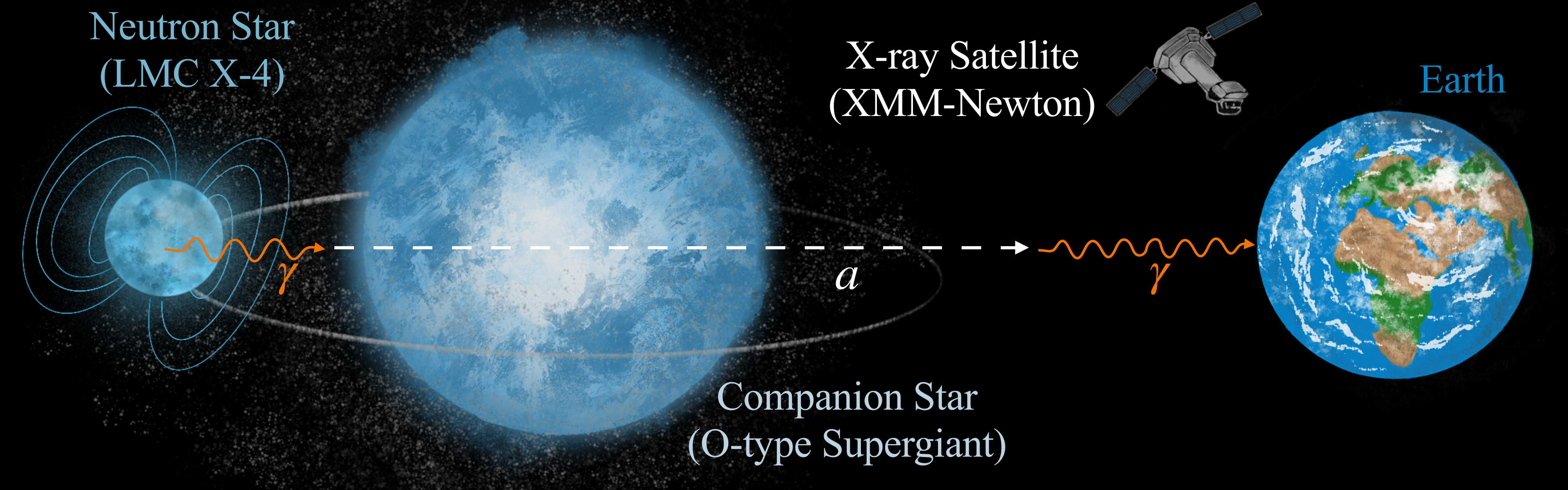}
    \caption{Artist's rendition of the LMC X-4 binary system during an eclipse. Reproduced with permission from Mom Chatterjee.}
    \label{fig:eclipse}
\end{figure*}
 We illustrate the idea of neutron star eclipses as ALP laboratories in \cref{fig:eclipse}, where we show a schematic diagram of a particular binary system, LMC X-4 \cite{Shtykovsky:2016mcy,Jain:2024yxw,Naik:2003fe}, during an eclipse.
This work is organized as follows. 
In \cref{sec2,sec3}, we discuss the photon-to-ALP conversion probability in the vicinity of the neutron star and the ALP-to-photon conversion probability in the interstellar medium, respectively. We also explain why LMC X-4 is the most suitable system for our study. In \cref{sec4}, using photon data from this system, we present constraints on the ALP-photon coupling strength $g_{a\gamma}$ as a function of ALP mass $m_a$. In \cref{sec5}, we conclude.

\smallskip
\noindent
\section{Photon to ALP transition near binary systems}
\label{sec2}
Neutron stars are bright sources of X-rays \cite{Paul:2017hqz,Rigoselli:2025opc,Rigoselli:2024fuq}, and in order to utilize the eclipse for constraining $g_{a\gamma}$ (see again \cref{fig:eclipse}), we require neutron star originated X-rays, traveling along the line of sight toward space observatories, to partially convert to ALPs in the region between the surface of the neutron star and its companion star, namely before reaching the `barrier'.

The two-level system featuring ALP and photon states can be described by the following effective Hamiltonian:
\begin{align}
\mathcal{H}_{\text{eff}}=-\begin{pmatrix}
\Delta_{\gamma} & \Delta_{a\gamma} \\
\Delta_{a\gamma} & \Delta_a
\end{pmatrix}\,,
\label{eq:ham}
\end{align}
where
\begin{align}
\Delta_{\gamma}&=-\frac{m_\text{eff}^2}{2 E_\gamma}\,,   &
\Delta_a&=-\frac{m_a^2}{2 E_\gamma}\,,  & 
\Delta_{a\gamma}&=\frac{1}{2} g_{a\gamma} |\vec{B}_T|\,.
\label{eq:elements}
\end{align}
Here, $E_\gamma$ is the photon energy, $\vec{B}_T$ is the component of the magnetic field orthogonal to the propagation direction, and $m_\text{eff}$ is the effective photon mass in the medium, which reads~\cite{Dobrynina:2014qba}
\begin{align}
m_\text{eff}^2(r) = \frac{4 \pi \alpha}{m_e} n_e(r) -\frac{88\, \alpha^2 E_\gamma^2}{270 \, m_e^4} B(r)^2 \,.
\label{eq:eff}
\end{align}
Here, $m_e$ is the electron mass and $\alpha$ is the fine-structure constant. We indicate with $r$ the spatial dependence of the electron number density in the medium, $n_e$, and the magnetic field $B$. We treat the magnetic field as a dipole,  
$B(r)=B^{(0)} \, (r/10 \, \text{km})^{-3}$, and also take $n_e(r)=n_e^{(0)}\, (r/10 \, \text{km})^{-3}$ \cite{Chattopadhyay:2023nuq}, where $n_e^{(0)}$ and $B^{(0)}$ are the values at the surface of the neutron star. 

\begin{figure}[b!]
	\centering
	\includegraphics[width=\linewidth]{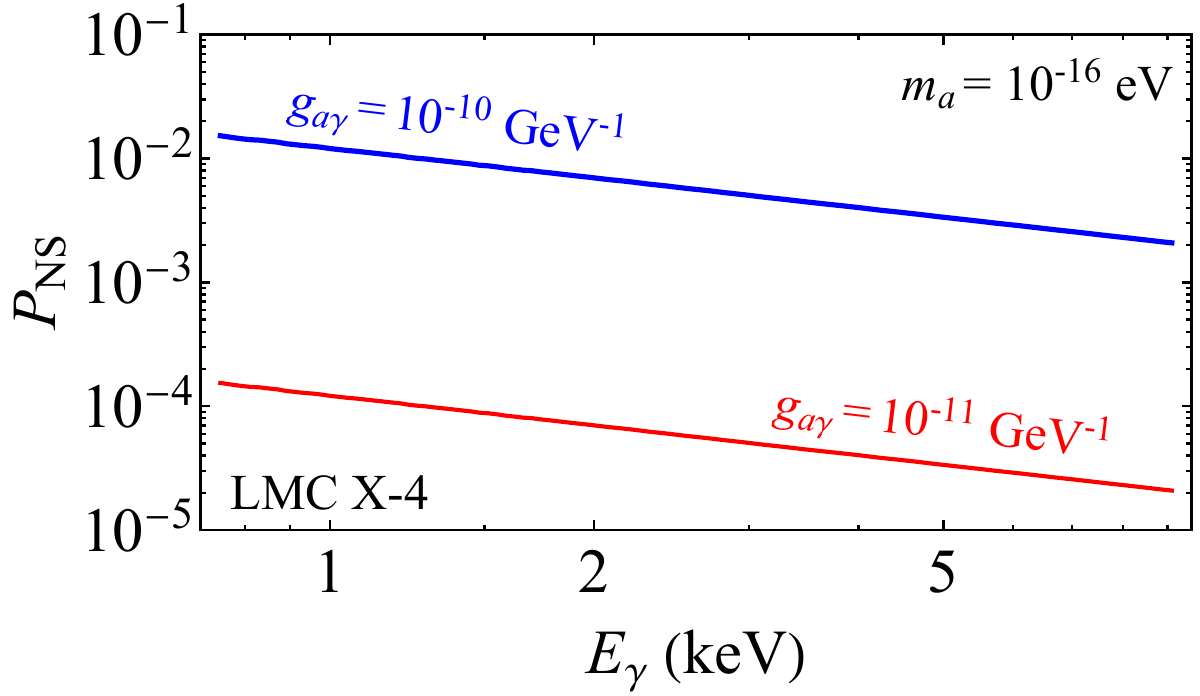}
	\caption{The photon-to-ALP transition probability, $P_\text{NS}$, near the neutron star in the LMC X-4 system for a fixed ALP mass and two values of $g_{a\gamma}$.}
	\label{fig:P1}
\end{figure}

Let us now discuss which eclipsing binary system is the most suitable for our study, as its identification will allow us to fix $n_e^{(0)}$ and $B^{(0)}$. High-mass X-ray binaries typically have magnetic fields several orders of magnitude stronger than those of low-mass systems, and higher values of $B^{(0)}$ enhance photon-to-ALP conversion near the neutron star. Among observed eclipsing high-mass X-ray binaries, a longer propagation distance would lead to higher ALP-to-photon transition probability in the interstellar medium. Further, systems where the observed decrease in luminosity during the eclipse is large, compared to the out-of-eclipse flux, would also lead to stronger results. Based on these criteria, we found the LMC X-4 binary system to be the most promising candidate.
For this system, the total magnetic field at the surface of the LMC X-4 neutron star is $B^{(0)} = 3 \times 10^{13}$ G \cite{Shtykovsky:2016mcy,Moon:2002zd,Rikame:2022jem}; we fix the transverse component of the magnetic field to $1.5 \times 10^{13}$ G, with the same power law dependence. For $n_e^{(0)}$, we take a benchmark value of $10^{14} \, \text{cm}^{-3}$; our results only mildly depend on $n_e^{(0)}$. The same holds even for deviation from the $r^{-3}$ dependence 
of $n_e$.

Armed with values for $n_e^{(0)}$ and $B^{(0)}$, we can quantify the amount of ALPs produced from  photons in the vicinity of the neutron star. We solve the Schr{\"o}dinger equation with $\mathcal{H}_{\text{eff}}$ from \cref{eq:ham}:  
\begin{align}  
i \frac{d}{dr} \begin{pmatrix}  
\gamma(r) \\  
a(r)  
\end{pmatrix} = \mathcal{H}_\text{eff} \begin{pmatrix}  
\gamma(r) \\  
a(r)  
\end{pmatrix}\,,  
\label{eq:diff}
\end{align}  
where $|\gamma(r)|^2$ and $|a(r)|^2$ represent the fractions of photons and ALPs at a given distance from the surface of the neutron star (the initial conditions at the neutron star surface are those of a pure photon state). 
For the LMC X-4 system, the photon-to-ALP conversion probability $P_{\text{NS}}$ is shown in \cref{fig:P1} for two values of $g_{a\gamma}$, for an ALP mass of $m_a = 10^{-16}$ eV. The photon-to-ALP transition probability depends primarily on the strength of the magnetic field around a neutron star, as well as the size of the neutron star itself, which regulates how fast the magnetic field will fall as photons travel outward from the surface of the star. We observe that the conversion probability $P_{\text{NS}}$ drops for $m_a \gtrsim 10^{-4}$ eV.

In our calculation, we neglect the ALP flux produced thermally inside the neutron star. Such a contribution will increase the total number of ALPs passing through the companion star, and thus neglecting it makes our analysis more conservative. Furthermore, we found that $a \leftrightarrow \gamma$ conversion inside the companion star, or in its magnetic field, is negligible and can be ignored.

\smallskip
\noindent
\section{ALP to photon transition in the interstellar medium}
\label{sec3}
Following ALP production in the magnetosphere of the neutron star during the eclipse, these hidden particles will partially convert back to photons en route to Earth. There are several magnetized regions where this process could occur: $(i)$ near the companion star, $(ii)$ in the interstellar medium of the Large Magellanic Cloud (LMC) where the eclipsing binary system LMC X-4 is located, $(iii)$ in the intergalactic medium between the LMC and the Milky Way, and $(iv)$ in the interstellar medium of the Milky Way. We have estimated that $(i)$ yields negligible contributions, and we conservatively do not take into account $(ii)$ and $(iii)$. 

\begin{figure}
    \centering
    \includegraphics[width=\linewidth]{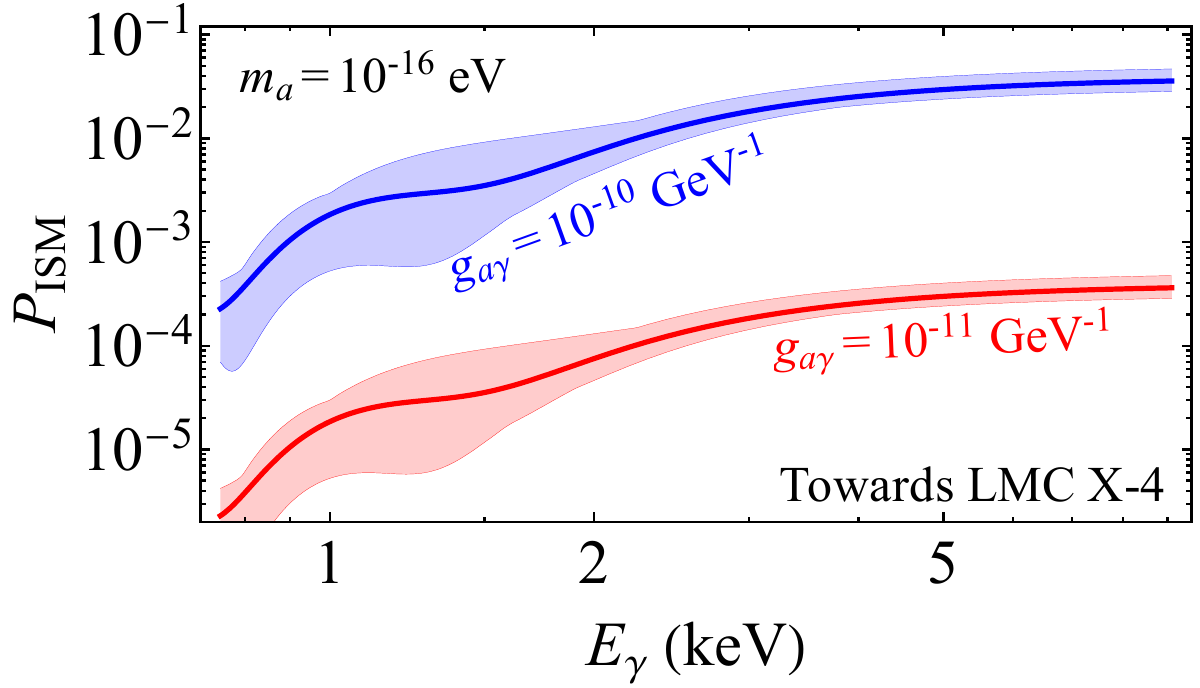}
    \caption{The ALP-to-photon transition probability, $P_{\text{ISM}}$, in the magnetic field of the Milky Way's interstellar medium for a fixed ALP mass and two values of $g_{a\gamma}$. The boundaries of the shaded regions correspond to the 
minimum and maximum $P_{\text{ISM}}$ values obtained using available models from \cite{Unger:2023lob} while the lines represent the mean values.}
    \label{fig:P2}
\end{figure}

For $(iv)$, knowledge of the galactic magnetic field and electron density is required, and we model these quantities following \cite{Unger:2023lob} and \cite{ne}, respectively. Specifically, for the magnetic field in the galaxy, we consider all eight realizations of UF23 model \cite{Unger:2023lob}, and for the electron density we use the YMW16 model \cite{ne}.
We compute ALP-to-photon transition probability in the interstellar medium of the Milky Way, $P_{\text{ISM}}$, by solving again \cref{eq:diff} (in this case $\mathcal{H}_{\text{eff}}$ contains $B$ and $n_e$ corresponding to the interstellar medium, varying  as a function of the distance along the line-of-sight)  
in the two flavor approximation. Conservatively, we only consider ALP-to-photon conversion in the last $10$ kpc of the traversed path. In \cref{fig:P2}, we show $P_{\text{ISM}}$ as a function of photon energy for two representative values of $g_{a\gamma}$. The eight different realizations of the magnetic field model yield different results for $P_{\text{ISM}}$, as indicated by the shaded regions. The mean value of $P_{\text{ISM}}$ is shown by the highlighted blue (red) line for $g_{a\gamma} = 10^{-10} \,\text{GeV}^{-1}$ ($g_{a\gamma} = 10^{-11} \,\text{GeV}^{-1}$).  We use the mean value of $P_{\text{ISM}}$ for obtaining the constraint on $g_{a\gamma}$ from the LMC X-4 eclipsing binary system.

We can also compare the results from \cref{fig:P2} with a simple analytical estimate. Magnetic field and electron density in the Milky Way can be approximated as $B_T \sim \text{few} \times 10^{-1} \mu$G and $n_e \sim \text{few} \times 10^{-3} \,\text{cm}^{-3}$ \cite{Jansson:2012pc,Unger:2023lob,ne}. With these values, terms from \cref{eq:elements} are approximately 
\begin{align}
\Delta_\gamma &\approx 1.1 \, \bigg(\frac{n_e}{10^{-2} \, \text{cm}^{-3}}\bigg) \, \left(\frac{ E_\gamma }{1\,\text{keV}}\right)^{-1} \, \text{kpc}^{-1}\,, & \nonumber \\
\Delta_{a\gamma}&\approx 0.15\, \left(\frac{g_{a\gamma}}{10^{-10} \, \text{GeV}^{-1}}\right) \left(\frac{B_T}{1\, \mu\text{G}} \right) \, \text{kpc}^{-1}\,,
\end{align}
with $\Delta_a\approx 0$ for $m_a\to 0$. For constant values of $n_e$ and $B_T$, in the limit of $\Delta_{\gamma}>\Delta_{a\gamma}\gg \Delta_a$, the analytical formula for $P_{\text{ISM}}$ simply reads $P_{\text{ISM}}= 4 (\Delta_{a\gamma}^2/\Delta_{\gamma}^2) \sin^2(\Delta_{\gamma} L/2)$. At galactic-scale distances the second term averages to $1/2$, yielding $P_{\text{ISM}}=2 (\Delta_{a\gamma}^2/\Delta_{\gamma}^2)$. For $E_\gamma=1$ keV and $g_{a\gamma}=10^{-10} \, \text{GeV}^{-1}$, value of $P_{\text{ISM}}\sim 10^{-2}$ is obtained, falling within the ballpark of our full numerical calculation, obtained for varying values of electron density (YMW16 model~\cite{ne}) and magnetic field strength (8 different realizations of the UF23 model~\cite{Unger:2023lob}).

\smallskip
\noindent
\section{Constraining the ALP-photon coupling with the eclipsing binary system LMC X-4}
\label{sec4}
\begin{figure}[t!]
    \centering
    \includegraphics[width=\linewidth]{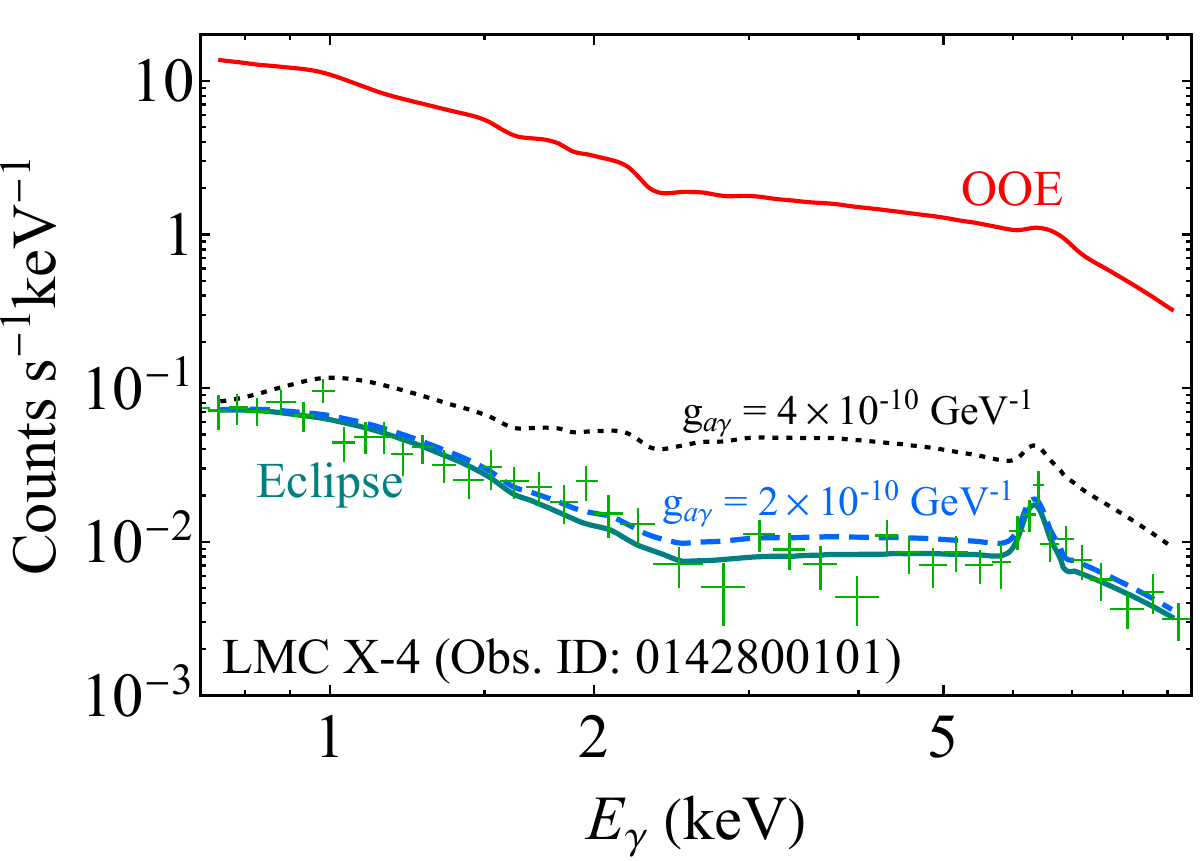}
    \caption{The photon fluxes observed by XMM-Newton (observation ID 0142800101) during the eclipse (green) and out-of-eclipse (red). We also present the contribution from the photon-ALP-photon process (blue dashed, and black dotted) for two different values of $g_{a\gamma}$ and $m_a=10^{-16}$ eV.
}
\label{fig:R}
\end{figure}

The total number of photons observed during the eclipse must exceed the number of photons produced through the photon-ALP-photon process. The contribution of the 
photon-ALP-photon process is equal to $F_{\text{OOE}} \, P_{\text{NS}} \, P_{\text{ISM}}$, where $F_{\text{OOE}}$ is the out-of-eclipse flux. This must be smaller than the observed flux during the eclipse
\begin{align}
F_{\text{eclipse}} \gtrsim F_{\text{OOE}}\, P_{\text{NS}}\, P_{\text{ISM}} \,.
\end{align}
The most conservative constraint would be obtained by solving the equation 
\begin{align}
P_{\text{NS}} P_{\text{ISM}} = F_{\text{eclipse}}/F_{\text{OOE}} \equiv 1/R\,,
\label{eq:es}
\end{align}
where $R$ is defined as the ratio of fluxes out of and during eclipse. Typical values of $R$ are $\mathcal{O}(10^2-10^3)$; see \cref{fig:R} where we show the out-of-eclipse flux (red) and the data associated with the eclipse (green), taken from Refs.~\cite{Aftab:2019ldh,Jain:2024yxw}. The dark green line is the fit provided by the authors of \cite{Aftab:2019ldh}. 

Let us first estimate using \cref{eq:es} the expected limit from the light-shining-through-wall mechanism ($\gamma \to\rm{ALP} \to \gamma$) during the eclipse of the LMC X-4 neutron star. From \cref{fig:P1,fig:P2}, we observe $P_{\text{NS}}\simeq 10^{-2}$ and $P_{\text{ISM}}\simeq 10^{-2}$ for $g_{a\gamma}=10^{-10}\, \text{GeV}^{-1}$ and $E_\gamma=3$ keV. Given the $g_{a\gamma}^2$ dependence for both conversion near the neutron star and in the interstellar medium, we expect to set a constraint at $g_{a\gamma} \simeq 2\times10^{-10}\, \text{GeV}^{-1}$, when the product $P_{\text{NS}} P_{\text{ISM}}$ reaches $10^{-3}$. This is precisely what we observe in \cref{fig:R}, where we also show the contribution of the photon-ALP-photon conversion probability, $P_{\text{NS}} P_{\text{ISM}}$, for $g_{a\gamma} \simeq 2\times10^{-10}\, \text{GeV}^{-1}$ and $g_{a\gamma} \simeq 4\times10^{-10}\, \text{GeV}^{-1}$. While the former (dashed blue) appears marginally constrained, the latter (dotted black) is visibly disfavored. Note that, since $P_{\text{NS}} P_{\text{ISM}}$ is maximum in $E_\gamma \approx (3-6)$ keV range, the energy dependence of photon-ALP-photon contribution is markedly different from the astrophysical flux during the eclipse (see again difference between spectra with and without ALP contribution in \cref{fig:R}).

We go beyond the above estimates to calculate LMC X-4 constraints on the ALP-photon coupling. Specifically, we compare the sum of the ALP contribution and the best-fit astrophysical photon flux during the eclipse (provided in Ref.~\cite{Aftab:2019ldh}) against the observed eclipse data from XMM-Newton (observation ID 0142800101) \cite{Aftab:2019ldh,Jain:2024yxw}. We define the corresponding chi-square value, for each choice of ALP mass $(m_a)$ and ALP-photon coupling $(g_{a\gamma})$, as
\begin{align}
	\chi^2_{\rm ALP} \equiv \chi^2_{\rm ALP+ecl.} -\chi^2_{\rm ecl.} \; ,
\label{eq:master}
\end{align}
where we abbreviate eclipse as $\rm{ecl.}$, for compact notation.
We define $\chi^2_{\rm ecl.} $ by summing over all the relevant energy bins, as
\begin{equation}
	\chi^2_{\rm ecl.} =\sum_{i} \Big(\widetilde F_{\rm ecl.}(E_i) - F_{\rm ecl.} (E_i) \Big)^2 \Big/ \Big(\delta \! F_{\rm ecl.} (E_i) \Big)^2 \; ,
\end{equation}
where $\widetilde F_{\rm ecl.}(E_i)$ denotes the best-fit average value of the flux during eclipse for the $i$-th energy bin, obtained by integrating the fit function over the energy-bin. The quantity $ \delta \! F_{\rm ecl.} (E_i)$ denotes the standard deviation of the experimental data corresponding to the energy bin $E_i$, taken from Fig. 2 of~\cite{Aftab:2019ldh}. 

We calculate $\chi^2_{\rm ALP+ecl.} $, for different values of $m_a$ and $g_{a\gamma}$, as
\begin{align}
	& \chi^2_{\rm ALP+ecl.}  \nonumber\\
	& \quad = \sum_{i} \Big(\widetilde F_{\rm ALP+ecl.}(E_i) - F_{\rm ecl.} (E_i) \Big)^2 \Big/ \Big(\delta \! F_{\rm ecl.} (E_i) \Big)^2 \; ,
\end{align}
where the flux $\widetilde F_{\rm ALP+ecl.}(E_i)$, for an energy bin $E_i \equiv (E_i^{l}, \, E_i^{r})$, is given by
\begin{align}
	& \widetilde F_{\rm ALP+ecl.}(E_i) = \widetilde F_{\rm ecl.}(E_i) \nonumber \\
	& \quad + \frac{ 1}{E_i^{r} -E_i^{l}} \int_{E_i^{l}}^{E_i^{r}} dE \, F_{\rm OOE} (E) \, P_{\rm NS} (E) \, P_{\rm ISM} (E) \, .
\end{align}
Note that both $P_{\rm NS} $ and $P_{\rm ISM}$ are also functions of $m_a$ and $g_{a\gamma}$.

Using the process outlined above, for each value of $m_a$, we vary the ALP-photon coupling $g_{a\gamma}$ and calculate $\chi^2_{\rm ALP}$. We find that the 90\% C.L. bound on the ALP-photon coupling, for $m_a \lesssim 10^{-12}$ eV, is at $g_{a\gamma} \approx 1.44 \times 10^{-10}$ GeV$^{-1}$.

\begin{figure}
    \centering
    \includegraphics[width=\linewidth]{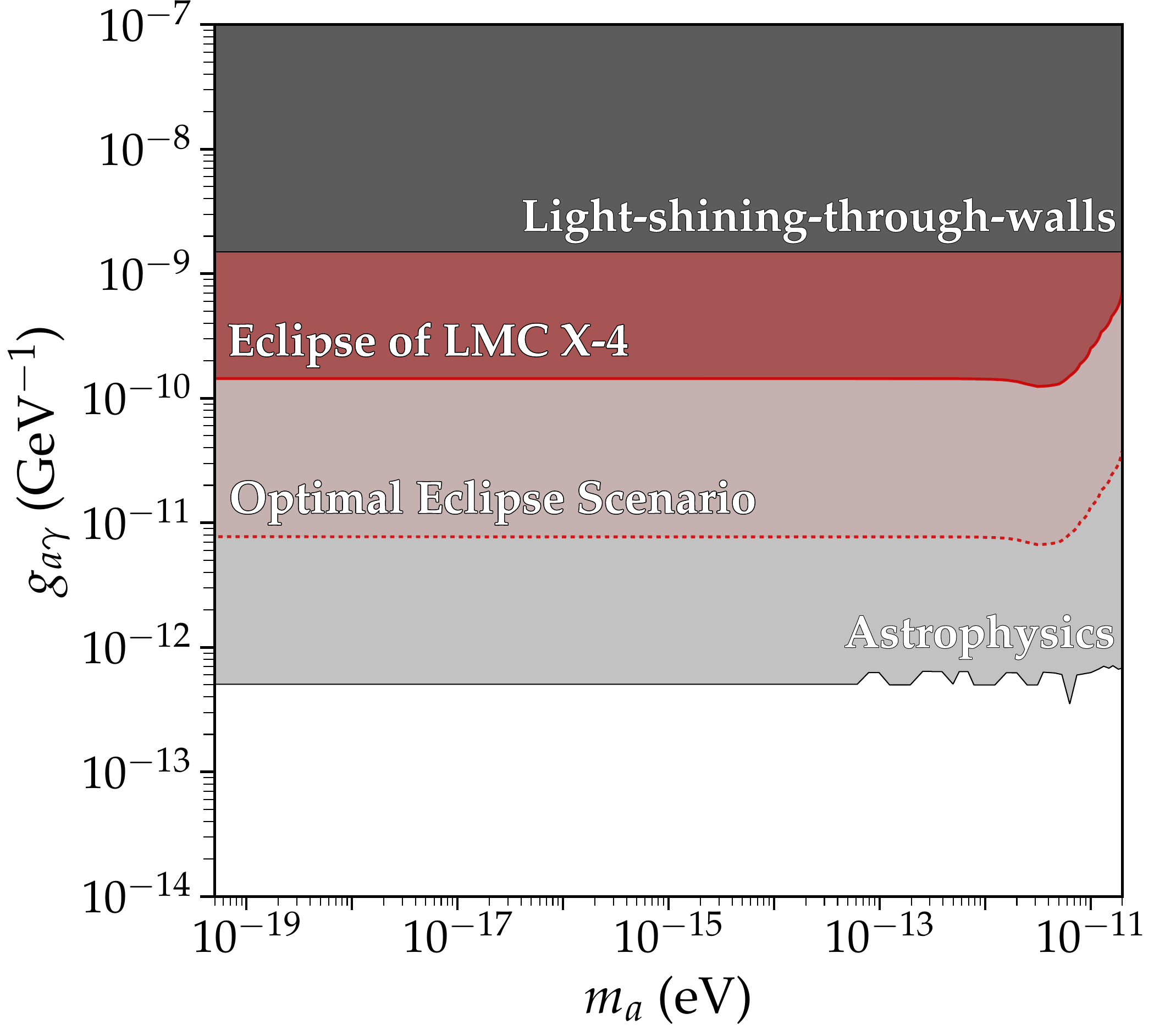}
    \caption{The constraint from LMC X-4 (red), reaching $g_{a\gamma} \approx 1.44\times 10^{-10}\, \text{GeV}^{-1}$ (at 90\% C.L.), is presented in the $m_a$--$g_{a\gamma}$ parameter space and compared with constraints from laboratory and astrophysics (gray). We also show the optimal eclipse scenario (light red) reaching $g_{a\gamma} \approx 8\times 10^{-12}\, \text{GeV}^{-1}$.}
    \vspace{-5pt}
\label{fig:money}
\end{figure}

In \cref{fig:money}, we present this constraint in the $m_a$--$g_{a\gamma}$ parameter space and compare it to the existing probes.  The limit from the eclipsing binary system LMC X-4 (red) is stronger by approximately one order of magnitude in $g_{a\gamma}$ compared to the first result from ALPS-II~\cite{ALPSII:2025eri} (dark gray) (see also previous generation of light-shining-through-walls experiments, e.g., ALPS I, OSQAR, and CROWS \cite{OSQAR:2007oyv,ALPS:2009des,Betz:2013dza}). Our limit is mostly flat in $g_{a\gamma}$ apart from a dip around $m_a\simeq 10^{-11}$ eV that stems from the  resonance $m_a^2=m_\text{eff}^2$ in the interstellar medium which enhances the sensitivity. For larger values of $m_a$ the limit sharply fades away. Our constraint remains weaker in comparison to the leading astrophysical limits \cite{Wouters:2013hua,Marsh:2017yvc,Meyer:2020vzy,Dessert:2020lil,Reynes:2021bpe,Ning:2024eky} (light gray), which benefit from $g_{a\gamma}^2$ dependence from $\gamma \leftrightarrow $ ALP, rather than photon reappearance ($g_{a\gamma}^4$ dependence from $\gamma\to \text{ALP} \to\gamma$) that is characteristic of the light-shining-through-walls concept.  Note however, that the same $g_{a\gamma}^4$ dependence also makes our constraints extremely robust, as even a small upward shift in the value of $g_{a\gamma}$ would results in a strongly modified $\gamma \to \rm{ALP} \to \gamma$ transition probability which cannot be accommodated.

 In \cref{fig:money}, we also show the eclipse scenario dubbed ``optimal,'' (light red) which is more competitive to some of the most sensitive probes.
For this optimal case, we consider how improvements across multiple fronts can strengthen the constraint on the ALP-photon coupling to $g_{a\gamma} \lesssim 10^{-11} \, \text{GeV}^{-1}$. In particular, we consider the following: $(i)$~better understanding of the eclipsing system, combined with longer observation time, may constrain the contribution of the photon-ALP-photon process to at most a few percent of the total observed flux during the eclipse (the current constraints allow the contribution of $\gamma \to \rm{ALP} \to \gamma$ process  approximately at $(10-20)$\%  for LMC X-4 during eclipse), $(ii)$~an eclipsing neutron star system located towards the Galactic Center would be ideal due to the stronger magnetic fields of $\sim$ few $\mu$G, which could lead to an $\sim \mathcal{O}(10)$ enhancement in $P_{\text{ISM}}$ (this enhancement is absent for LMC X-4, given that the line of sight does not pass through the Galactic Center), $(iii)$~a stronger magnetic field of the candidate neutron star (e.g., an eclipsing magnetar) could increase $P_{\text{NS}}$ by an order of magnitude, and $(iv)$~future observations may achieve unprecedented angular resolution for such a system, potentially enhancing the out-of-eclipse to eclipse flux ratio by $\sim \mathcal{O}(100)$. While $(i)$ depends on future improvements in our understanding of eclipsing neutron star systems, we have estimated the effects of $(ii)$ and $(iii)$ to be viable given the identification of an ideal eclipsing system. Regarding $(iv)$, observations of radial profiles of low-mass X-ray binary systems~\cite{Aftab:2023yqk} already demonstrate an improvement of $\sim\mathcal{O} (10-20)$ in the out-of-eclipse to eclipse flux ratio achieved with $\sim$arcsecond angular resolution (see Fig.~1 in~\cite{Aftab:2023yqk}). A future probe of eclipsing high-mass X-ray binary systems, with sub-arcsecond angular resolution, may then be able to distinguish the contribution of photons scattered toward Earth during the eclipse, leading to an $\mathcal{O}(100)$ improvement in the flux ratio. Combined, the discussed improvements $(i)$-$(iv)$ would strengthen the constraint on the ALP-photon coupling strength $g_{a\gamma}$ by a factor of $\sim 20$, resulting in the projected sensitivity for the optimal eclipse scenario shown in~\cref{fig:money}.

\smallskip
\noindent
\section{Summary}
\label{sec5}
In this work, we have presented a novel method for searching for axions and ALPs based on the light-shining-through-walls concept. We utilized eclipsing binary systems composed of a neutron star, which is a bright source of X-rays with very high magnetic field strengths, and a companion star, through which ALPs produced via conversion in the neutron star's magnetosphere can pass during the eclipse. The ALPs then partially reconvert into photons in the interstellar medium on their way to Earth, and the resulting X-rays are detectable by space observatories such as XMM-Newton and NuSTAR. We have found that, at present, the most suitable eclipsing binary system for ALP searches is LMC X-4, for which we derived constraints at the level of $g_{a\gamma}\simeq 10^{-10}\, \text{GeV}^{-1}$, surpassing current laboratory constraints from light-shining-through-walls experiments by one order of magnitude in $g_{a\gamma}$, for ALPs with masses below $10^{-11}$ eV. We have also discussed possible improvements that, in an optimal scenario, could enhance the sensitivity by more than an order of magnitude. Overall, in this work, we have proposed an exciting new probe that complements and extends current efforts to search for elusive, light pseudoscalar bosons.

\smallskip
\noindent
\section*{Acknowledgments}
We would like to thank Kaladi Babu, Bryce Cyr, and Benjamin Lehmann for useful discussions. We also thank Mom Chatterjee for illustrating the schematic diagram (\cref{fig:eclipse}). The work of VB is supported by the United States Department of Energy Grant No. DE-SC0025477.

\bibliographystyle{JHEP}
\bibliography{refs}

\end{document}